\documentclass[aps,prb,reprint,amsmath,amssymb %,groupedaddress,frontmatterverbose
]{revtex4-1}
\usepackage{graphicx}% Include figure files
\usepackage[colorlinks=true,allcolors=blue]{hyperref}%
\usepackage{tikz}
\newcommand{\yr}{Y}
\begin{document}

%\title{\href{run:D:/PhD/Thesis/My Papers/Paper 2/Draft/mypaper.tex}{A Tunable Spherical Graphene Spaser}}
\title{A Tunable Spherical Graphene Spaser}

\author{Sadreddin Behjati Ardakani}
\email[]{behjati@ee.sharif.ir}
\affiliation{Department of Electrical Engineering, Sharif University of Technology, Azadi Avenue, Tehran, Iran}

\author{Rahim Faez}
\email[]{faez@sharif.ir}
\affiliation{Department of Electrical Engineering, Sharif University of Technology, Azadi Avenue, Tehran, Iran}

\date{\today}

\begin{abstract}
In this work, a new structure is suggested for spasing. The presented spaser is made up of a graphene nanosphere, which supports localized surface plasmon modes, and a quantum dot array, acting as a gain medium. The gain medium is pumped by an external laser source. Since all the plasmons are carried on a graphene platform, the structure features coherent surface plasmons with high confinement and large life time. All the structure is analyzed theoretically using full quantum mechanical description. The main advantage of the proposed spaser is the simple tuning capability of it by changing graphene's Fermi level which is performed by either chemical doping in the manufacturing time or electrostatic gating. We suggest utilizing the proposed spaser for exciting coherent, long range surface plasmons on a graphene sheet. The near field of the spaser couples to the surface plasmons on graphene sheet and compensates the large momentum mismatch between surface plasmons and photons.
\end{abstract}

%\pacs{73.20.Mf, 42.50.Nn}
%\keywords{Plasmonics, Graphene, Spaser}

\maketitle

\section{\label{sec:int}Introduction}
A SPASER (Suarface Plasmon Amplification by Stimulated Emission of Radiation) is a plasmon nanolaser. It does not suffer from the diffraction limit of light, so it can work in deep subwavelength dimensions. A spaser emits intense coherent Surface Plasmons (SPs) instaed of photons. Spasing action was first introduced by Stockman and Bergman in 2003.\cite{bergman2003surface} The authors showed that equations of motion of such a system have a stationary solution, in absence of any input, under some specific conditions. They showed that this phenomenon has only a full quantum mechanical description. Since introduction of the spaser, some researchers around the world have been working on different approaches for realizing and analyzing the spaser. In 2009, Noginov \textit{et al.} demonstrated an experimental spaser which was made of an aqueous solution of gold nanoparticles surrounded by dye-doped silica shell as the gain medium.\cite{noginov2009demonstration} In 2010, Stockman proposed a SP amplifier using the spaser idea.\cite{stockman2010spaser} In 2013, Dorfman \textit{et al.} developed the theory of spaser for three level systems, in contrast to the previous two level model.\cite{dorfman2013quantum} The authors showed that a three level system can acquire the spasing condition much easier than two level ones. Since the discovery of spasing phenomenon, many research papers have published which explored different aspects of a spaser.\cite{andrianov2011forced, khurgin2012injection, li2013electric, parfenyev2014quantum, rupasinghe2014spaser, jayasekara2015multimode, totero2016energy}

A spaser, the same as a laser, requires at least two media to work, an active or gain medium and one for supporting plasmonic modes. SP modes can exist on interface between two media which one of them has a negative dielectric constant. Negative permitivity is a characteristic feature of metals below their plasma frequencies. That is why the majority of works, in the field of plasmonics, utilizes metals. High Joule losses in metals cause SPs to have short propagation lengths or short life times. The alternative candidates for supporting SP modes are 2D materials which by modifying the boundary conditions allow SP modes to exist. Graphene is the most famous 2D material, which is proved to support SP modes with an order of magnitude better propagation length, confinement, and life time.\cite{jablan2009plasmonics}

Graphene is a material which is formed by 2D arrangement of carbon atoms in a honeycomb lattice bonded by strength sp$^2$ hybridized $\sigma$ bonds. This material has a unique linear dispersion around Dirac points, which endows graphene some extraordinary features. Graphene's electrons which lie near Dirac points behave like massless Dirac fermions. Due to unique properties of SPs in graphene, some researchers proposed using of graphene in designing the spaser.\cite{apalkov2014proposed,jayasekara2015multimode,berman2013graphene}

In this paper, a spherical graphene structure is proposed for spasing. In this structure, graphene has the role of Localized Surface Plasmon (LSP) supporter and an array of Quantum Dots (QDs) is used as the gain medium. This structure is analyzed quantum mechanically using two level description. In quantizing the Hamiltonian of the system, we use a more general approach, instead of the popular one,\cite{jayasekara2015multimode,rupasinghe2014spaser,chang2007strong,rukhlenko2009spontaneous,apalkov2014proposed} that is widely used in the literature. In the method used in the present work, the whole Hamiltonian is written down, including the kinetic energy of electrons in graphene as well as potential one. For deriving the kinetic energy of 2D electrons in graphene, we define an effective mass which is compatible with the graphene's conductivity.

One of the main issues, concerning SPs, is the method for exciting them. SPs on metal films or graphene sheets have a large momentum mismatch with photons of plane wave light. Exciting SPs accomplishes by some elegant methods, such as prism coupling, grating, near field excitation, and so on.\cite{maier2007plasmonics} One of the applications of our work could be utilizing the near field of the spaser for exciting SPs on graphene sheets or metal films.

The best description for spasing action is gained by full quantum mechanical treatment. In this picture both the field and matter are quantized. The Hamiltonian of the whole system consists of three parts, \mbox{$H=H_\mathbf{LSP}+H_\mathrm{g}+H_\mathrm{I}$}, where the terms on the RHS from left to right are LSP, gain, and interaction Hamiltonians, respectively. In order to quantize the Hamiltonian, we need the orthogonal potential modes of the structure and dipole moment of the gain medium. Based on this discussion, the paper is organized as follows. In section \ref{sec:structure}, the proposed structure is introduced. Section \ref{sec:LSPHam}, deals with LSP Hamiltonian and its quantization. Section \ref{sec:gainHam} is devoted to the investigation of the active medium and deriving its dipole moment in order to write its Hamiltonian. In section \ref{sec:IntHam}, the interaction Hamiltonian is assessed and the spasing condition is derived.

\section{\label{sec:structure}The main structure}
\begin{figure}[tb]
	\centering
	\includegraphics[width=\linewidth]{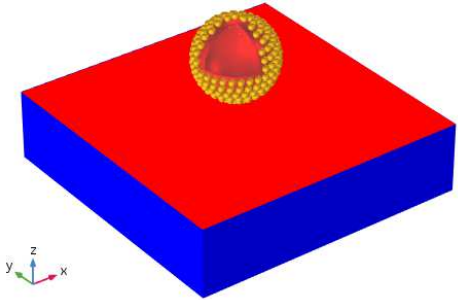}
	\caption{\label{fig:struct}(Color online) The proposed structure which is used in this paper. The red color illustrates graphene and the yellow color represents QDs. The picture is not drawn to scale. A section of sphere is removed for ease of illustration.}
\end{figure}

The proposed structure consists of a shell of spherical graphene, with radius $a$, surrounded by an array of QDs, each of which has a radius equals to $b$, where $b\ll a$. The graphene sphere together with QDs are playing the role of a spaser. The spaser stands on top of a graphene sheet which sits on a substrate. The structure is shown in Fig.~\ref{fig:struct}.

The whole system is composed of two main parts, the spaser nanosphere and the graphene sheet which will guide the plasmons out of the spaser. These parts are investigated, separately, in the ongoing sections.

The nanosphere spaser, which is going to be a generator of LSPs, is made up of two subsystems, QD array and the spherical graphene shell. QD array is a gain medium, transferring energy to the LSPs, and graphene nanosphere supports LSP modes. The spaser is embedded in a matrix with dielectric constant $\epsilon_\mathrm{ra}$. Dielectric constant of the inner part of the nanosphere is assumed to be $\epsilon_\mathrm{rp}$.

Although the analyses in this paper are general, but for the purpose of numerical calculations, the specific materials with parameters shown in Table~\ref{tab:param} \begin{table}[bt]
	\caption{\label{tab:param}Physical parameters of materials which are used in this paper. All the alloys are chosen such that to be lattice matched with each other. Dielectric constants of ternary alloys are calculated from interpolation method.\cite{chuang2009physics,vurgaftman2001band}}
	\begin{ruledtabular}
		\begin{tabular}{lccc}
			Material & $\epsilon_\mathrm{r}$\footnote{Dielectric constant,} & $m^*/m_0$\footnote{Electron's effective mass in units of $m_0$, where $m_0$ is electron's mass which is equal to $9.1\times10^{-31}\,\mathrm{Kg}$,} & $E_\mathrm{g}$\footnote{Energy gap in $e\mathrm{V}$,}\\
			\colrule
%			$\mathrm{InP}$ & 12.56\footnote{Data is extracted from Ref.~\onlinecite{chuang2009physics}.} & 0.077\footnotemark[4] & 1.344\footnotemark[4]\\
			$\mathrm{Al}_{0.48}\mathrm{In}_{0.52}\mathrm{As}$ & 12.46\footnote{Data is calculated by interpolation method,} & 0.075\footnote{Data is extracted from Ref.~\onlinecite{chuang2009physics}.} & 1.450\footnotemark[5]\\
			$\mathrm{Ga}_{0.47}\mathrm{In}_{0.53}\mathrm{As}$ & 13.60\footnotemark[4] & 0.041\footnotemark[5] & 0.750\footnotemark[5]\\
%			$\mathrm{SiO}_2$ & 3.9 & ---
		\end{tabular}
	\end{ruledtabular}
\end{table}
are used. Each QD is formed by $\mathrm{Al}_{0.48}\mathrm{In}_{0.52}\mathrm{As}/\mathrm{Ga}_{0.47}\mathrm{In}_{0.53}\mathrm{As}$ heterostructure (barrier/well). The alloy fractions are chosen such that the well and barrier are lattice matched with each other. The matrix and substrate are both made of $\mathrm{Al}_{0.48}\mathrm{In}_{0.52}\mathrm{As}$.

Pumping is accomplished by illuminating the system by an external laser source. The pumping energy excites the electrons in QDs which then couple to LSPs. The external laser source, directly, has not any effect on graphene sheet because the momentum mismatch between photons and SPs avoids any coupling. After generating LSPs by spaser, these quasiparticles could outcouple utilizing near field excitation. 

In the rest of the paper, we analyze the nanosphere spaser alone because the existence of graphene sheet can be assessed perturbatively.

\section{\label{sec:LSPHam}LSP Hamiltonian}
Before beginning the quantization of the LSP Hamiltonian, The orthogonal potential modes of the structure should be derived. So the next subsection is dedicated to extracting the LSP modes of the graphene nanosphere and in the last subsection, $H_\mathrm{LSP}$ is quantized.
 
\subsection{\label{subsec:modes}LSP modes of graphene nanosphere}
For extracting the LSP modes of the structure, precisely, the full wave nature of the field should be considered. However, according to the extra confine character of LSPs, the modes could be derived by utilizing quasi-electrostatic approximation. So Laplace equation for the electrostatic potential, $\Phi$, should be solved, $\nabla^2\Phi=0$. 
%Because of the spherical symmetry of the structure, the spherical coordinate system is used throughout the paper. The notation $r$, $\theta$, and $\phi$ is used for radius, inclination, and azimuth angle, respectively. The unit vectors are indicated by adding a hat symbol above the associated direction.

Exploiting the symmetry of the structure, potential modes can be written in the following form,
\begin{equation}
\Phi_{lm}(r,\theta,\phi)=\left\lbrace\begin{array}{ll}
A_{lm}r^l\yr_l^m(\theta,\phi) &r\le a,\\
B_{lm}r^{-(l+1)}\yr_l^m(\theta,\phi) &r>a,
\end{array}\right.
\end{equation}
where $A_{lm}$ and $B_{lm}$ are unknown coefficients of $lm$'th mode to be determined and $\yr_l^m$s are spherical harmonic functions. 
%All the angular dependencies are integrated into spherical harmonic functions. 
%The first lowest order modes are sketched in Fig.~\ref{fig:modes}\begin{figure}[tb]
%	\centering
%	\begin{tikzpicture}[x=1cm,y=1cm]
%	\draw (-.8,0) -- (7.3,0) (0,.8) -- (0,-5.8);
%	\node[inner sep=0pt] (pic) at (3.9,-3) {\includegraphics[width=.85\linewidth, trim=65 40 40 31, clip=true]{modes.eps}};
%	\draw (-.5,-.8) node {$l=1$} (-.5,-3) node {$l=2$} (-.5,-5.2) node {$l=3$} (.7,.5) node {$m=0$} (2.75,.5) node {$m=1$} (4.75,.5) node {$m=2$} (6.75,.5) node {$m=3$};
%	\end{tikzpicture}
%	%	\includegraphics[width=\linewidth, trim=40 30 30 20, clip=true]{modes.eps}	
%	\caption{\label{fig:modes}(Color online) This figure shows the angular dependencies of the first four lowest order modes. This figure excludes the obvious monopole mode, $l=0$, which has no angular dependency.}
%\end{figure}. The monopole mode, $l=0$, is not included in this figure because of obvious angle independency.
From now on, for simplicity, the angular dependencies of spherical harmonics are not written, explicitly. Applying the continuity of potential on graphene's interface leads to
\begin{equation}
\frac{A_{lm}}{B_{lm}}=a^{-(2l+1)}.\label{eq:coefrel1}
\end{equation}
The second independent boundary condition relates the discontinuity of electric field across the boundary to the surface charge density, $\hat{\mathbf{n}}\cdot(\mathbf{D}_{lm}^\mathrm{a}-\mathbf{D}_{lm}^\mathrm{p})=\rho_{lm}^\mathrm{s}$,where $\hat{\mathbf{n}}$, $\mathbf{D}_{lm}$, and $\rho^\mathrm{s}_{lm}$ are unit vector normal to the interface, electric displacement, and surface charge density of $lm$'s mode, respectively. The superscript indices $\mathrm{a}$ and $\mathrm{p}$ indicate two sides of interface, ambient and nanosphere, respectively. Surface charge density could be derived by using the current continuity equation on the graphene in frequency domain, $\nabla_\mathrm{T}\cdot\mathbf{J}_{lm}^\mathrm{s}-i\omega_{lm}\rho_{lm}^\mathrm{s}=0$, where $\exp(-i\omega t)$ convention is used for time dependence and $\nabla_\mathrm{T}$ indicates tangential Del operator and $\mathbf{J}_{lm}^\mathrm{s}$ is surface current density on graphene. If this relation is combined with the Ohm's law, $\mathbf{J}_{lm}^\mathrm{s}=\sigma_\mathrm{s}\mathbf{E}_{lm}^\mathrm{T}$, where $\mathrm{T}$ denotes tangential component and $\sigma_\mathrm{s}$ is the surface conductivity of nanosphere, the following formula is obtained,
\begin{equation}
\rho_{lm}^\mathrm{s}=\frac{1}{i\omega_{lm}}\nabla_\mathrm{T}\cdot\sigma_\mathrm{s}\mathbf{E}_{lm}^\mathrm{T},\label{eq:surfchrel}
\end{equation}
and the electric field is
\begin{eqnarray}
\mathbf{E}_{lm} &=&\left\lbrace\begin{array}{ll}
A_{lm}r^{(l-1)}l &r<a\\
-B_{lm}r^{-(l+2)}(l+1) &r>a
\end{array}\right\rbrace\hat{r}\yr_l^m\nonumber\\
&+&\left[\frac{\partial}{\partial\theta}\hat{\theta}+im\csc\theta\hat{\phi}\right]\yr_l^m.
\end{eqnarray}
Using the above relation for electric field and substituting in Eq.~(\ref{eq:surfchrel}), $\rho^\mathrm{s}_{lm}$ is derived,
\begin{equation}
\rho_{lm}^\mathrm{s}=\frac{A_{lm}a^{(l-2)}\sigma_\mathrm{s}}{i\omega}\left[\cot\theta\frac{\partial}{\partial\theta}+\frac{\partial^2}{\partial\theta^2}-m^2\csc\theta\right]\yr_l^m. \label{eq:surfch}
\end{equation}
Substituting  Eq.~(\ref{eq:surfch}) into the normal electric field boundary condition yields the second relation between the coefficients,
\begin{equation}
\frac{A_{lm}}{B_{lm}}=\frac{\epsilon_\mathrm{ra}/l}{\sigma_\mathrm{s}/i\omega_{lm}\epsilon_0a-\epsilon_\mathrm{rp}/(l+1)}\times\frac{1}{a^{2l+1}},\label{eq:coefrel2}
\end{equation}
where $\epsilon_0$ is vacuum's permitivity. In deriving the above relation, the associated Legendre differential equation is utilized. Combining Eq.~(\ref{eq:coefrel1}) and Eq.~(\ref{eq:coefrel2}) leads to the following relation,
\begin{equation}
\frac{\sigma_\mathrm{s}(\omega_{lm})}{i\omega_{lm}\epsilon_0a}=\frac{\epsilon_\mathrm{rp}}{l+1}+\frac{\epsilon_\mathrm{ra}}{l}.\label{eq:disprel}
\end{equation}
This result resembles that of Ref.~\onlinecite{christensen2015localized} which is derived by different method using Mie theoty and could be a verification for our approach. 
%The result shows that the structure could not be precisely described by quasi-electrostatic approximation because it distinguishes between $\epsilon_\mathrm{rp}$ and $\epsilon_\mathrm{ra}$, in contrast to the electrostatic feature in which the important parameter is arithmetic average of dielectric constants, not each one's separately. However as the $l$ becomes larger and larger the approximation leads to more accurate results. For large enough $l$, Eq.~(\ref{eq:disprel}) takes the following form,
%\begin{equation}
%\frac{\sigma_\mathrm{s}(\omega_{lm})}{2i\omega_{lm}\epsilon_0a}=\frac{\bar{\epsilon}}{l},
%\end{equation}
%where $\bar{\epsilon}=0.5(\epsilon_\mathrm{rp}+\epsilon_\mathrm{ra})$. For the rest of this paper the quasi-electrostatic approximation is supposed to be sufficiently precise.

Equation (\ref{eq:disprel}) should be solved for unknown eigenfrequencies $\omega_{lm}$'s. Actually, Eq.~(\ref{eq:disprel}) is an implicit complex equation for complex variable $\Omega_{lm}=\omega_{lm}-i\gamma'_{lm}$, where $\gamma'_{lm}$ is $lm$'th mode's damping. This equation is decomposed to two independent real equations. The complexity of deriving eigenfrequencies could be reduced if the low loss nature of LSPs is considered ($\gamma'_{lm}\ll\omega_{lm}$). Having considered it, the system of complex equations collapses to the following real decoupled ones,
\begin{eqnarray}
\omega_{lm} &=& \frac{\sigma''_\mathrm{s}(\omega_{lm})}{\epsilon_0a[\epsilon_\mathrm{rp}/(l+1)+\epsilon_\mathrm{ra}/l]},\label{eq:modefreq}\\
\gamma'_{lm} &=& \frac{\omega_{lm}\sigma'_\mathrm{s}(\omega_{lm})}{\sigma''_\mathrm{s}(\omega_{lm})-\omega\,\partial/\partial\omega\,\sigma''_\mathrm{s}(\omega)|_{\omega=\omega_{lm}}},\label{eq:modedamp}
\end{eqnarray}
where we have assumed $\sigma_\mathrm{s}=\sigma'_\mathrm{s}+i\sigma''_\mathrm{s}$.
Eigenfrequencies and modes' dampings are derived using Eq.~(\ref{eq:modefreq}) and Eq.~(\ref{eq:modedamp}), respectively. Equation~(\ref{eq:modefreq}) can be solved using numerical root finding techniques.

A noteworthy result, deduced from Eq.~(\ref{eq:disprel}), is $2l+1$ degeneracy of potential modes, which is expected previously due to the symmetry considerations. For a fixed $l$, all the $\omega_{lm}$'s are the same. The quality factor of modes could be derived using $Q_{lm}=\omega_{lm}/2\gamma'_{lm}$.\cite{stockman2011nanoplasmonics}

Until now, no specific surface conductivity profile is assumed and then the results are general. But for further proceeding, the graphene's conductivity is accounted. Graphene conductivity is $\sigma_\mathrm{s}=\sigma_\mathrm{intra}+\sigma_\mathrm{inter}$, where intraband and interband conductivities are given by the following formulae,\cite{christensen2017classical}
\begin{equation}
\sigma_\mathrm{intra} =\frac{2e^2k_BT}{\pi\hbar^2}\frac{i}{\omega+i\tau^{-1}}\ln\left[2\cosh\left(\frac{E_\mathrm{F}}{2k_BT}\right)\right],
\end{equation}
and
\begin{eqnarray}
&&\!\!\!\!\!\!\sigma_\mathrm{inter}=\frac{e^2}{4\hbar}\times\nonumber\\
&&\!\!\!\!\!\!\left(\mathrm{H}(\omega/2)+\frac{4i(\omega+i\tau^{-1})}{\pi}\int_0^\infty\frac{[\mathrm{H}(\epsilon)-\mathrm{H}(\omega/2)]\,\mathrm{d}\epsilon}{(\omega+i\tau^{-1})^2-4\epsilon^2}\right),
\end{eqnarray}
where $e$, $k_B$, $\hbar$, $T$, $\omega$, $E_\mathrm{F}$, and $\tau\simeq0.4\mathrm{ps}$ \cite{apalkov2014proposed} are elementary charge, Boltzmann's and reduced Planck's constants, absolute temperature, angular frequency, Fermi energy, and electron's relaxation time, respectively. In the above relation, $H$ is defined as
\begin{equation}
\mathrm{H}(\epsilon)=\frac{\sinh(\hbar\epsilon/k_BT)}{\cosh(E_\mathrm{F}/k_BT)+\cosh(\hbar\epsilon/k_BT)}.
\end{equation}
It can be shown that for $\hbar\omega<2E_\mathrm{F}$ and \mbox{$\hbar\omega<\hbar\omega_\mathrm{oph}$}, where $\hbar\omega_\mathrm{oph}\simeq0.2\,\mathrm{eV}$ is optical phonon's energy, the graphene's conductivity is best approximated by Drude-like profile,\cite{jablan2009plasmonics}
\begin{equation}
\sigma_\mathrm{s}=\frac{e^2E_\mathrm{F}}{\pi\hbar^2}\frac{i}{\omega+i\tau^{-1}},\label{eq:drude}
%\sigma_\mathrm{inter} &=&\frac{e^2}{4\hbar}\left[\Theta(\hbar\omega-2\left|E_\mathrm{F}\right|)+\frac{i}{\pi}\ln\left|\frac{\hbar\omega-2\left|E_\mathrm{F}\right|}{\hbar\omega+2\left|E_\mathrm{F}\right|}\right|\right]\!.
\end{equation}
If the frequency of interest is in the range of validity of Drude approximation, the eigenfrequencies and quality factors can be derived analytically,
\begin{eqnarray}
\omega_{lm} &=& \left(\frac{e^2E_\mathrm{F}}{\pi\hbar^2\epsilon_0a(\epsilon_\mathrm{rp}/(l+1)+\epsilon_\mathrm{ra}/l)}-\frac{1}{4\tau^2}\right)^{1/2},\label{eq:apxomega}\\
\gamma'_{lm} &=& \frac{1}{2\tau},\label{eq:apxgamma}\\
Q_{lm} &=& \left(\frac{e^2E_\mathrm{F}\tau^2}{\pi\hbar^2\epsilon_0a(\epsilon_\mathrm{rp}/(l+1)+\epsilon_\mathrm{ra}/l)}-\frac{1}{4}\right)^{1/2}.\label{apxq}
\end{eqnarray}

\begin{figure*}[tb]
	\centering
	\includegraphics[width=.4\textwidth]{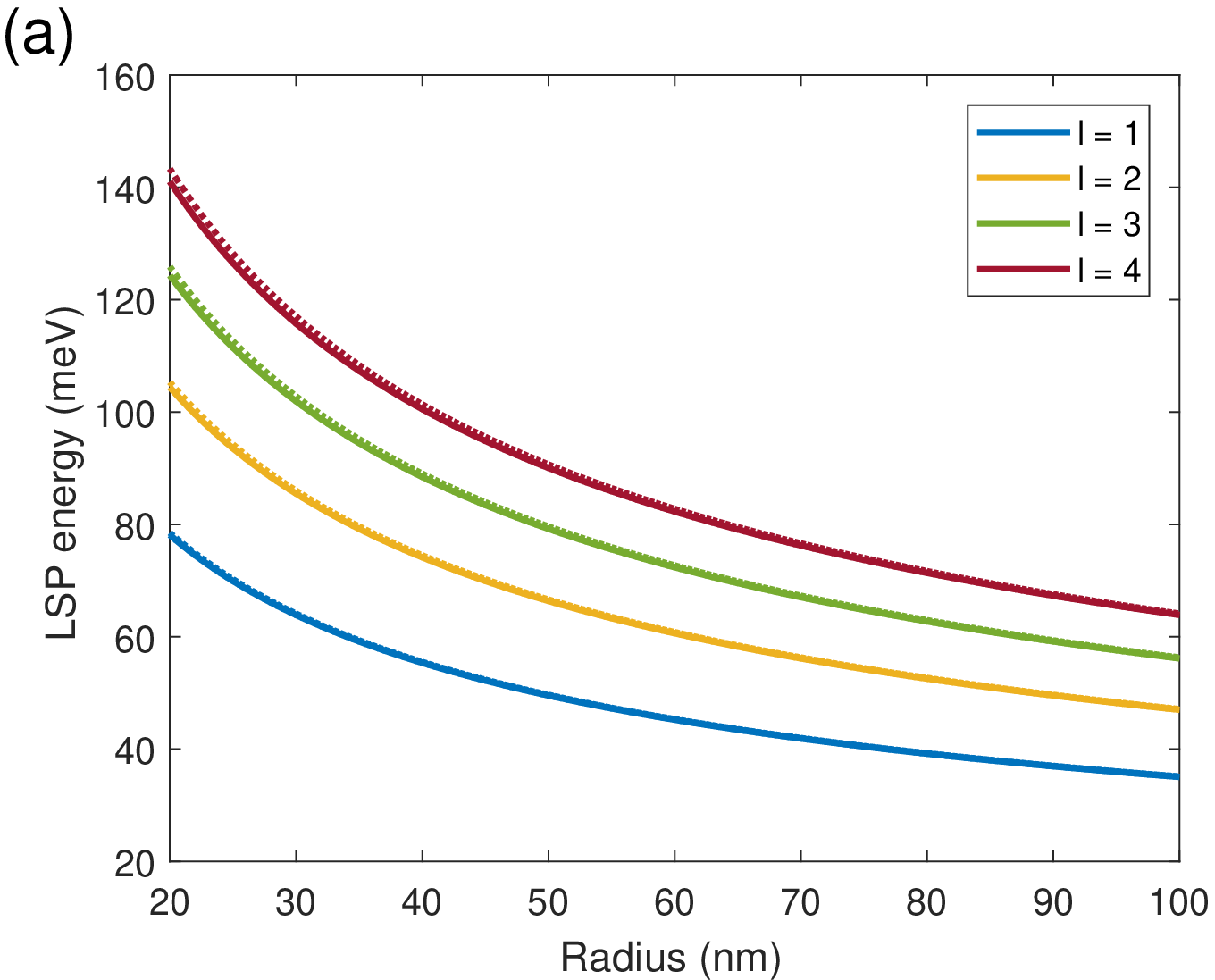}
	\includegraphics[width=.4\textwidth]{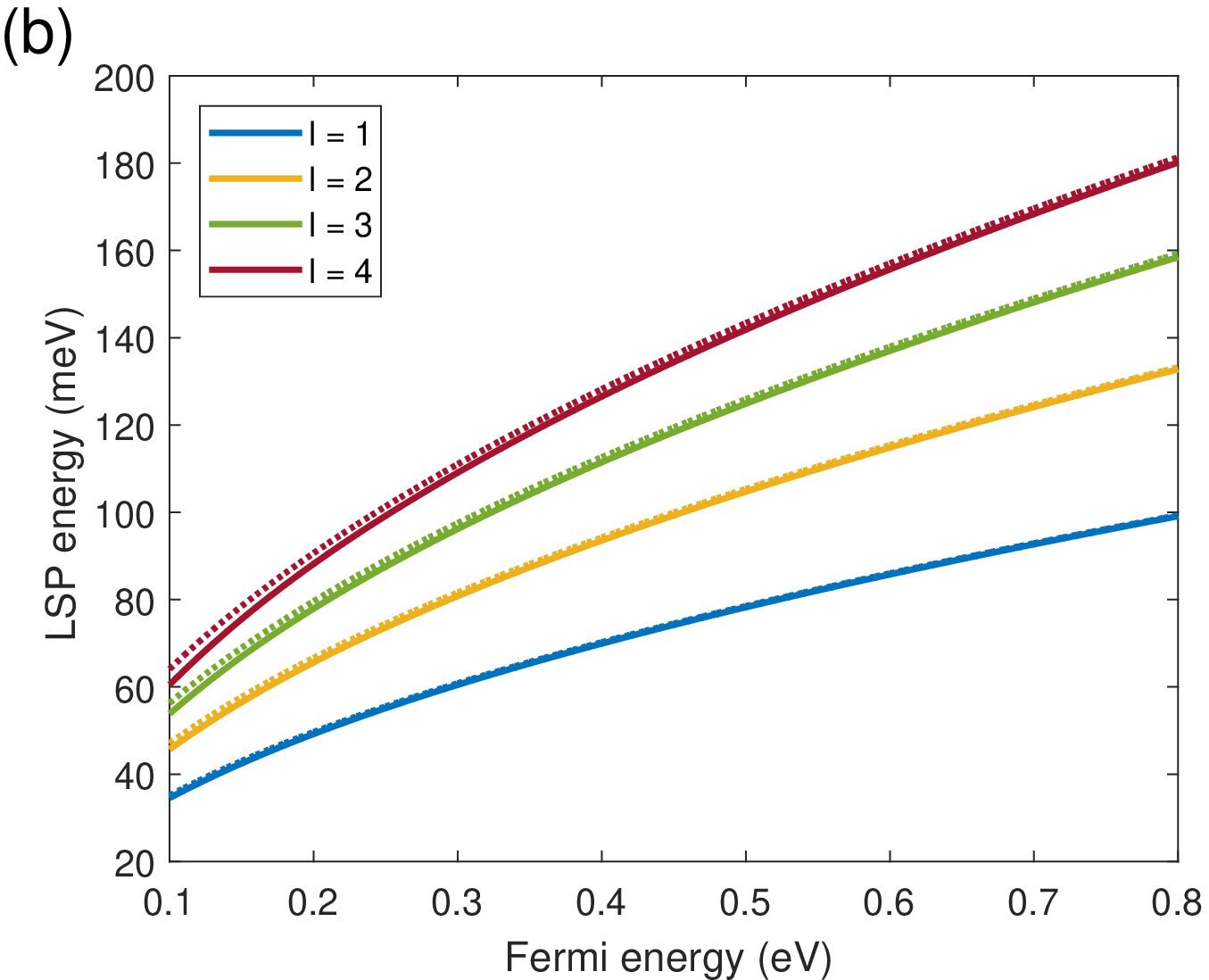}\\
	\includegraphics[width=.4\textwidth]{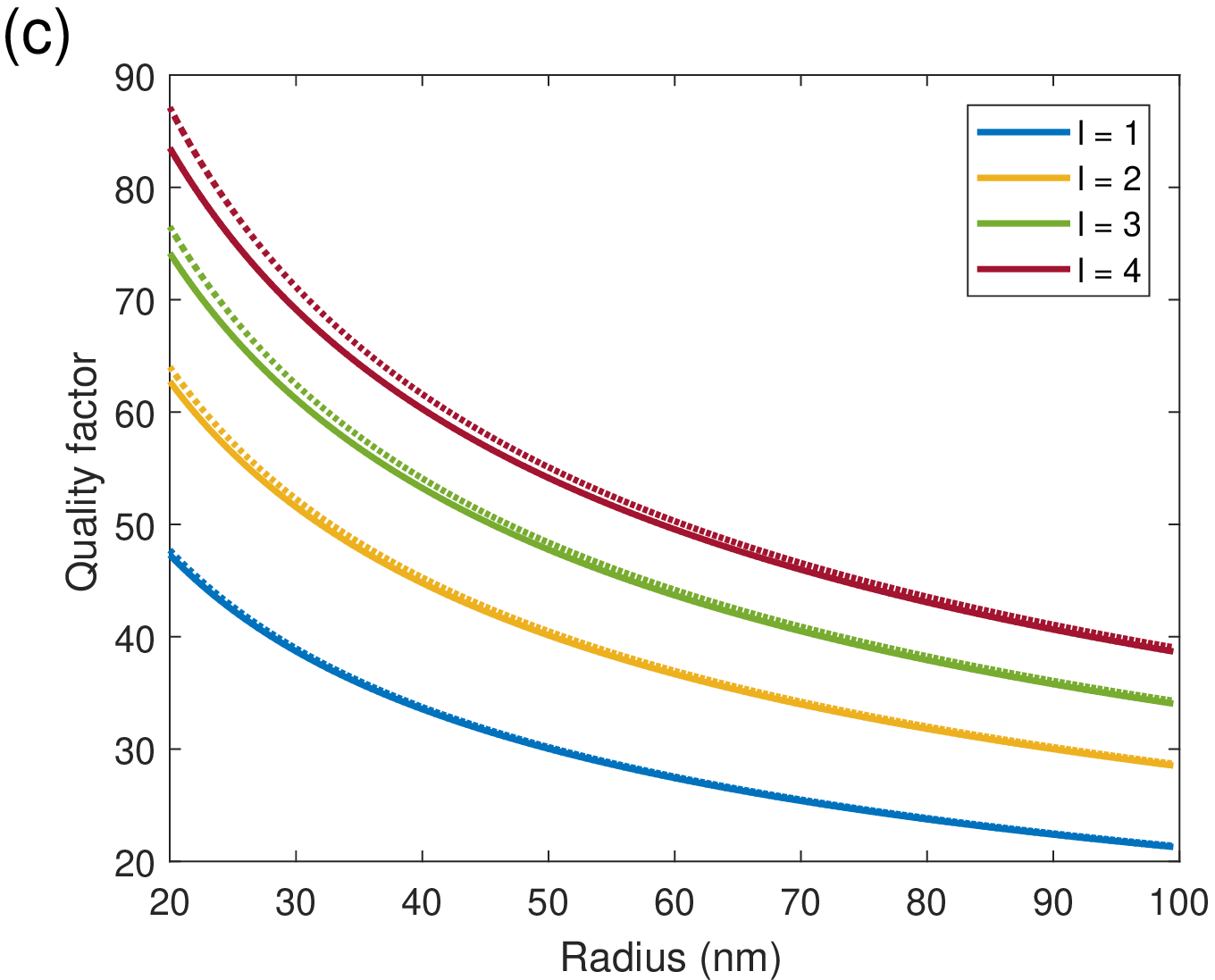}
	\includegraphics[width=.4\textwidth]{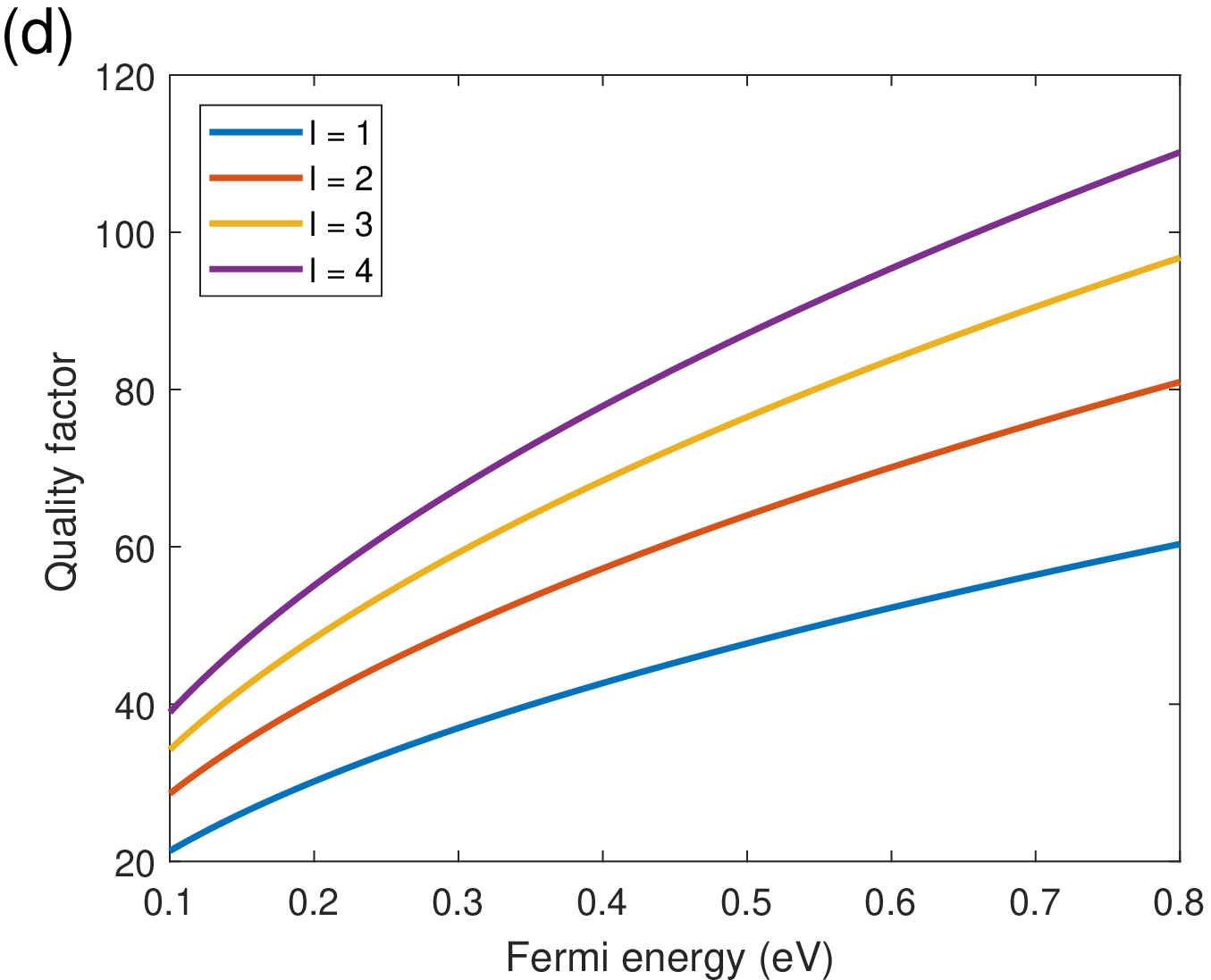}
	\caption{\label{fig:modefeatures}(Color online) (a) Energy of LSP versus graphene nanosphere's radius, (b) Energy of LSP against Fermi energy, (c) Quality factor as a function of graphene nanosphere's radius, and (d) Quality factor versus Fermi energy of the first four lowest order modes. (a) and (c) is sketched for Fermi energy of $0.4\,e\mathrm{V}$. In (b) and (d) the nanosphere's radius is assumed to be $25\,\mathrm{nm}$. All the parts are drawn assuming room temperature, $300\,\mathrm{K}$. In all the curves solid and dotted lines represent precise and approximate solutions, respectively. (d) is sketched only for approximate result.}
\end{figure*}

Figure~\ref{fig:modefeatures} shows various features of LSPs. Figure~\ref{fig:modefeatures}(a) and (b) sketch energy of a single LSP for various modes as a function of graphene nanosphere's radius and Fermi energy, respectively. Figure~\ref{fig:modefeatures}(c) and (d) illustrate quality factor of previous modes versus radius and Fermi energy, respectively. It is beneficial to compare the precise and approximate curves. It can be seen that for the shown range of radii and Fermi energies, the approximate results, Eqs.~(\ref{eq:apxomega}) and (\ref{apxq}), could estimate the actual values very well.

\begin{figure*}[tb]
\centering
\includegraphics[width=.4\textwidth]{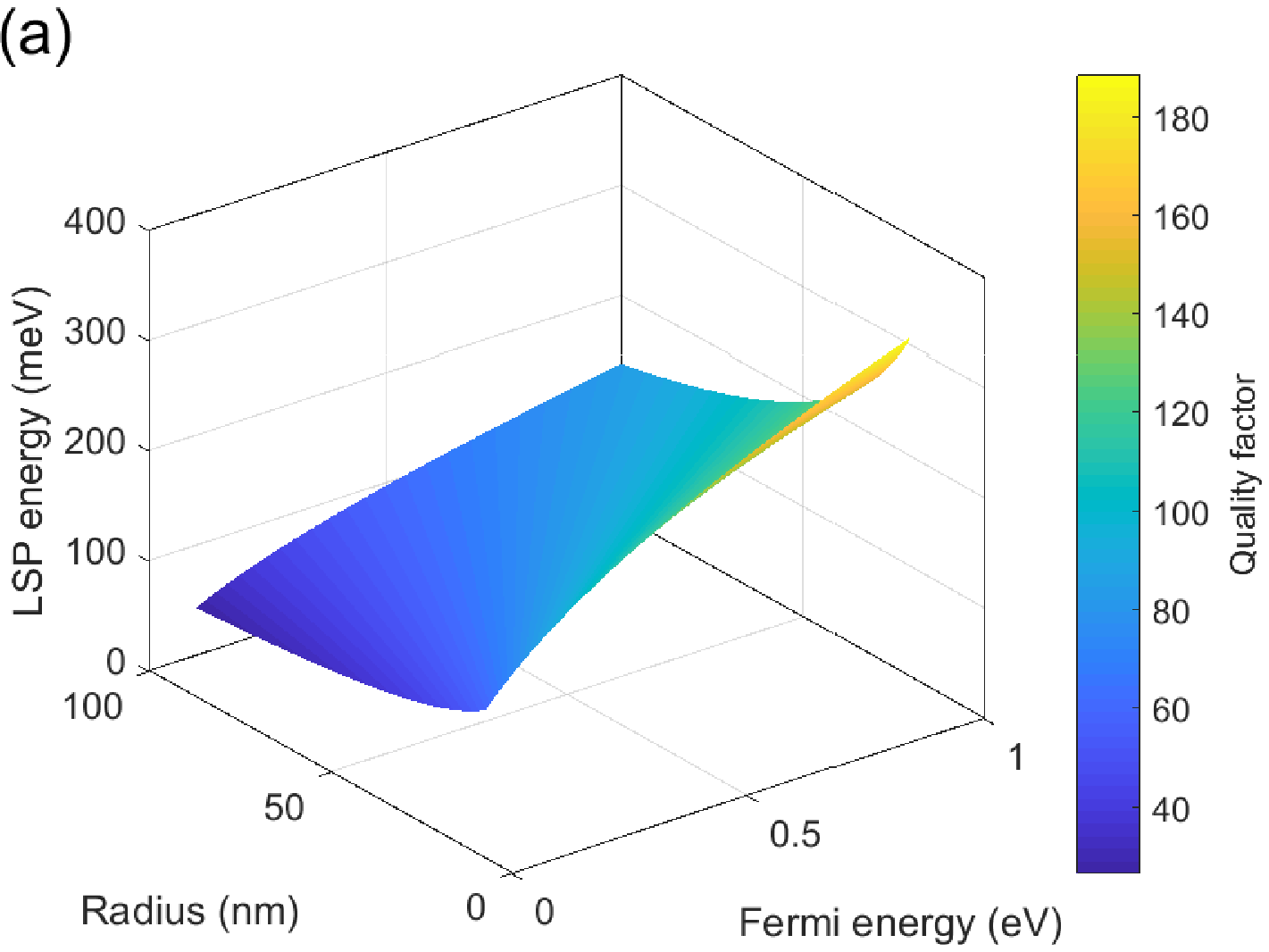}
\includegraphics[width=.4\textwidth]{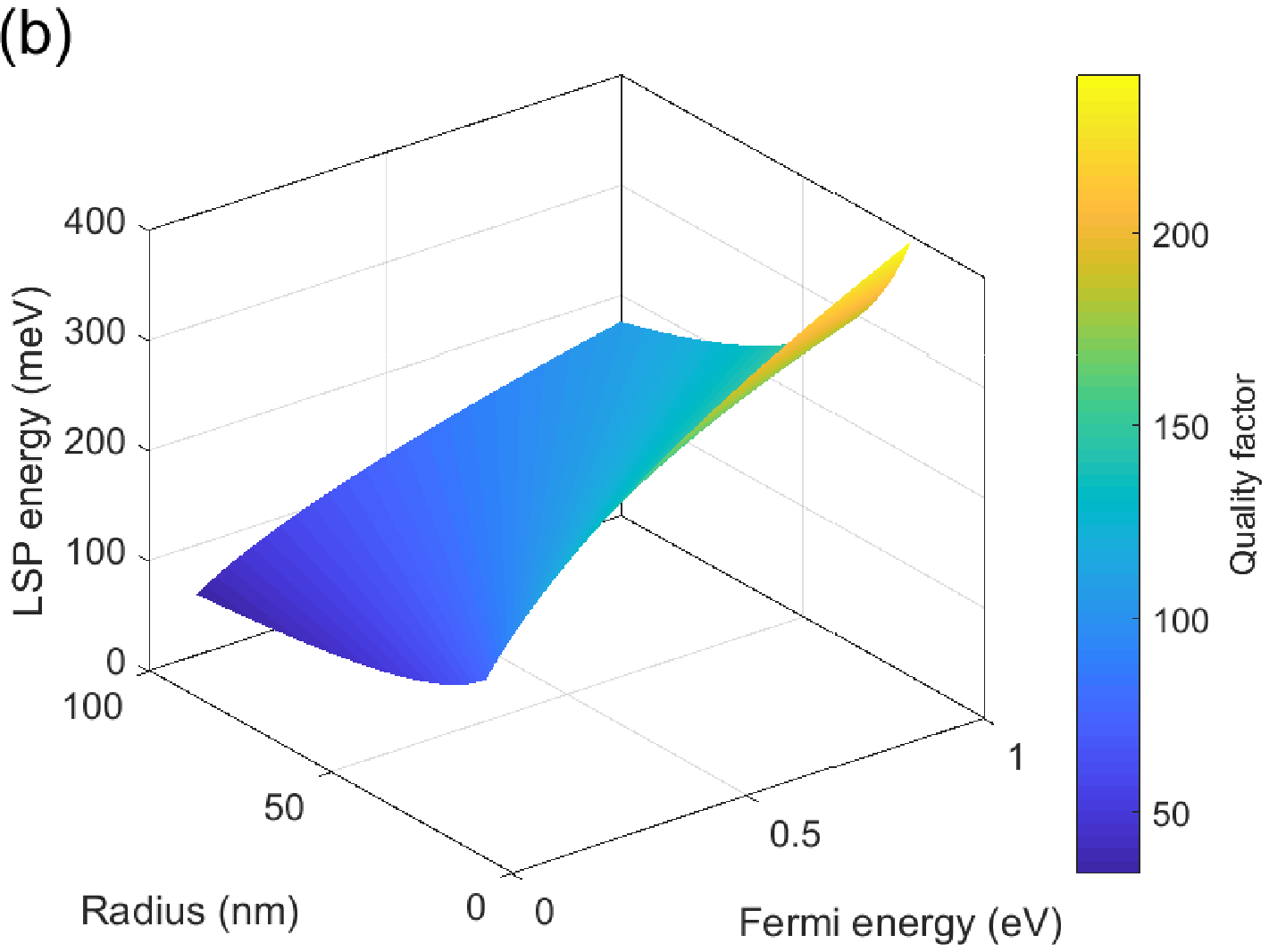}
\caption{\label{fig:mode2dfeatures}(Color online) LSP's energy and quality factor as a multivariable function of Fermi energy and nanosphere's radius for (a) dipole mode, and (b) quadrupole mode. In the figure height and color of surfaces represent LSP's energy and quality factor, respectively.}
\end{figure*}

Figure~\ref{fig:mode2dfeatures} represents the LSP energy and quality factor as a multivariable function of Fermi energy and nanosphere's radius simultaneously. Figure~\ref{fig:mode2dfeatures} (a) and (b) illustrate the aforementioned quantities for dipole and quadrupole mode, respectively.

\subsection{\label{subsec:LSPHamquan}Quantization of LSP Hamiltonian} 
LSP Hamiltonian is composed of two parts which are electrostatic and kinetic energy of electrons inside graphene nanosphere. The electrostatic Hamiltonian is due to the interaction between electric field and surface charges in graphene and the kinetic energy part is due to the motion of electrons. So it is reasonable to write $H_\mathrm{LSP}=H_\mathrm{kin}+H_\mathrm{es}$, where $H_\mathrm{kin}$ and $H_\mathrm{es}$ are kinetic and electrostatic Hamiltonians, respectively. The electrostatic energy can be found using the following well known relation,\cite{griffiths1962introduction}
\begin{equation}
H_\mathrm{es}=\frac{1}{2}\int_\mathrm{S}\rho_\mathrm{s}(\theta,\phi)\Phi(a,\theta,\phi)\,\mathrm{d}^2r,
\end{equation}
and the kinetic part is found using the following relation,
\begin{equation}
H_\mathrm{kin}=\frac{1}{2}n_{s0}m^*_e\int_\mathrm{S}\left|\mathbf{v}(\theta,\phi)\right|^2\,\mathrm{d}^2r,
\end{equation}
where $\rho_\mathrm{s}$, $n_{s0}$, and $\mathbf{v}$ are total surface charge density, equilibrium surface number density, and velocity of electrons, respectively, and integration is performed over the graphene's surface, $\mathrm{S}$. The above kinetic Hamiltonian resembles that of 3D electron gas. We suggest using of the same formulation for graphene, but with a modified electron effective mass, $m^*_e$, which is introduced using graphene's conductivity. We propose to find the effective mass by equating the Drude conductivity of graphene, Eq.~(\ref{eq:drude}), to the 3D electron gas one,\cite{ashcroft1976solid}
\begin{equation}
\sigma(\omega)=\frac{i\epsilon_0\omega_p^2}{\omega+i\gamma},\label{eq:sigmadr}
\end{equation}
where plasma frequency is defined by $\omega_p=e^2n_0/\epsilon_0m^*_e$ and $n_0$ is electron's number density. Doing such a way, the effective mass is found, $m^*_e=n_{s0}\pi\hbar^2/E_\mathrm{F}$.
%In the above relation $m^*_e$ is a custom defined effective mass equal to $m^*_e=n_{s0}\pi\hbar^2/E_\mathrm{F}$.

All the eigenmodes of the electrostatic potential are found in section~\ref{sec:LSPHam}. Because eigenvectors of a normal operator span the solution space, so the electric potential can be written as follows,
\begin{equation}
\Phi(r,\theta,\phi)=\sum_{lm}C_{lm}\Phi_{lm}(r,\theta,\phi)+\mathrm{c.c.},
\end{equation}
where $C_{lm}$'s are expansion coefficients and $\mathrm{c.c.}$ stands for complex conjugate of previous terms. In the above relation $\Phi_{lm}=\Phi_{lm}^++\Phi_{lm}^-$, where
\begin{eqnarray}
\Phi_{lm}^+ &=&\Theta(r-a)\left(\frac{r}{a}\right)^{-(l+1)}\yr_l^m(\theta,\phi),\\
\Phi_{lm}^- &=&\Theta(-r+a)\left(\frac{r}{a}\right)^l\yr_l^m(\theta,\phi).
\end{eqnarray}
Taking gradient of potential yields the electric field,
\begin{eqnarray}
\mathbf{E}_{lm}(r,\theta,\varphi)&=&\sum_{lm}C_{lm}\left\lbrace\begin{array}{ll}
l\frac{r^{l-1}}{a^l} &r<a\\
-(l+1)\frac{r^{-(l+2)}}{a^{-(l+1)}} &r>a
\end{array}\right\rbrace\hat{r}\yr_l^m\nonumber\\
&+&\left[\frac{\partial}{\partial\theta}\hat{\theta}+im\csc\theta\hat{\phi}\right]\yr_l^m.
\end{eqnarray}
Applying perpendicular boundary condition immediately leads to,
\begin{eqnarray}
\hspace{-5mm}\rho_\mathrm{s}(\theta,\phi)&=&\epsilon_\mathrm{ra}E_{r\mathrm{a}}-\epsilon_\mathrm{rp}E_{r\mathrm{p}}\nonumber\\
&=&\sum_{lm}C_{lm}\left[\epsilon_2\frac{l+1}{a}+\epsilon_1\frac{l}{a}\right]\yr_l^m(\theta,\phi)+\mathrm{c.c.}.
\end{eqnarray} 
The only quantity that should be derived is electron's velocity vector field in graphene. It can be found by using Newton's second law, $e\mathbf{E}_{lm}^\mathrm{T}=i\omega_{lm}m^*_e\mathbf{v}_{lm}$,
\begin{equation}
\mathbf{v}(\theta,\phi)=\sum_{lm}\frac{eC_{lm}}{i\omega_{lm}m^*_ea}\left[\frac{\partial}{\partial\theta}\hat{\theta}-\frac{im}{\sin\theta}\hat{\phi}\right]\yr_l^m(\theta,\phi)+\mathrm{c.c.}.
\end{equation}
Now, all of the variables requiring for extracting LSP Hamiltonian is provided. Using these relations and after some tedious algebra the following results are obtained,
\begin{eqnarray}
H_\mathrm{es} &=& \sum_{lm}\frac{l(l+1)\sigma_\mathrm{s}(\omega_{lm})}{2i\omega_{lm}}\left[C_{lm}^*C_{lm}+C_{lm}C_{lm}^*\right],\\
H_\mathrm{kin} &=& \sum_{lm}\frac{n_{s0}e^2l(l+1)}{2m^*_e|\omega_{lm}|^2}\left[C_{lm}^*C_{lm}+C_{lm}C_{lm}^*\right].
\end{eqnarray}
So the LSP Hamiltonian is derived,
\begin{equation}
H_\mathrm{LSP}=\sum_{lm}\frac{l(l+1)\sigma_\mathrm{s}(\omega_{lm})}{i\omega_{lm}}\left[C_{lm}^*C_{lm}+C_{lm}C_{lm}^*\right].
\end{equation}
If the low loss approximation for graphene could be assumed to be valid, the Hamiltonian is more simplified,
\begin{equation}
H_\mathrm{LSP}=\sum_{lm}\frac{l(l+1)\sigma''_\mathrm{s}(\omega_{lm})}{\omega_{lm}}\left[C_{lm}^*C_{lm}+C_{lm}C_{lm}^*\right].
\end{equation}
This relation becomes analogous to harmonic oscillator's Hamiltonian if the following modifications are performed,
\begin{eqnarray}
C_{lm} &\rightarrow& \gamma_{lm}\hat{a}_{lm},\\
C_{lm}^* &\rightarrow& \gamma_{lm}\hat{a}_{lm}^\dagger,
\end{eqnarray}
where
\begin{equation}
\gamma_{lm}^2=\frac{\hbar\omega_{lm}^2}{2l(l+1)\sigma''_\mathrm{s}(\omega_{lm})}.
\end{equation}
The annihilator and creator operators, $\hat{a}_{lm}$ and $\hat{a}^\dagger_{lm}$, satisfy the bosonic operator's algebra. By using these changes, the Hamiltonian recasts to an operator,
\begin{equation}
\hat{H}_\mathrm{LSP}=\sum_{lm}\frac{\hbar\omega_{lm}}{2}(\hat{a}_{lm}^\dagger \hat{a}_{lm}+\hat{a}_{lm}\hat{a}_{lm}^\dagger).
\end{equation}
So the electric field operator is
\begin{equation}
\hat{\mathbf{E}}=\sum_{lm}\gamma_{lm}(\mathbf{M}_{lm}^*\hat{a}_{lm}^\dagger+\mathbf{M}_{lm}\hat{a}_{lm}),
\end{equation}
where
\begin{eqnarray}
\mathbf{M}_{lm} &=&-\nabla\Phi_{lm}=\left\lbrace\begin{array}{ll}
l\frac{r^{l-1}}{a^l} &r<a\\
-(l+1)\frac{r^{-(l+2)}}{a^{-(l+1)}} &r>a
\end{array}\right\rbrace\hat{r}\yr_l^m\nonumber\\
&+&\left[\frac{\partial}{\partial\theta}\hat{\theta}+im\csc\theta\hat{\phi}\right]\yr_l^m.\label{eq:M}
\end{eqnarray}

In passing to the next subsection, we introduce some notation simplification. It is seen that quantities like $\omega_{lm}$, $\gamma'_{lm}$, $Q_{lm}$, and $\gamma_{lm}$ do not have any dependency on value of $m$. So for simplicity, we drop the $m$ index for only the aforementioned quantities and call them $\omega_l$, $\gamma'_l$, $Q_l$, and $\gamma_l$.

\section{\label{sec:gainHam}Active medium Hamiltonian}
\begin{figure}[tb]
	\centering
	\includegraphics[width=.15\textwidth]{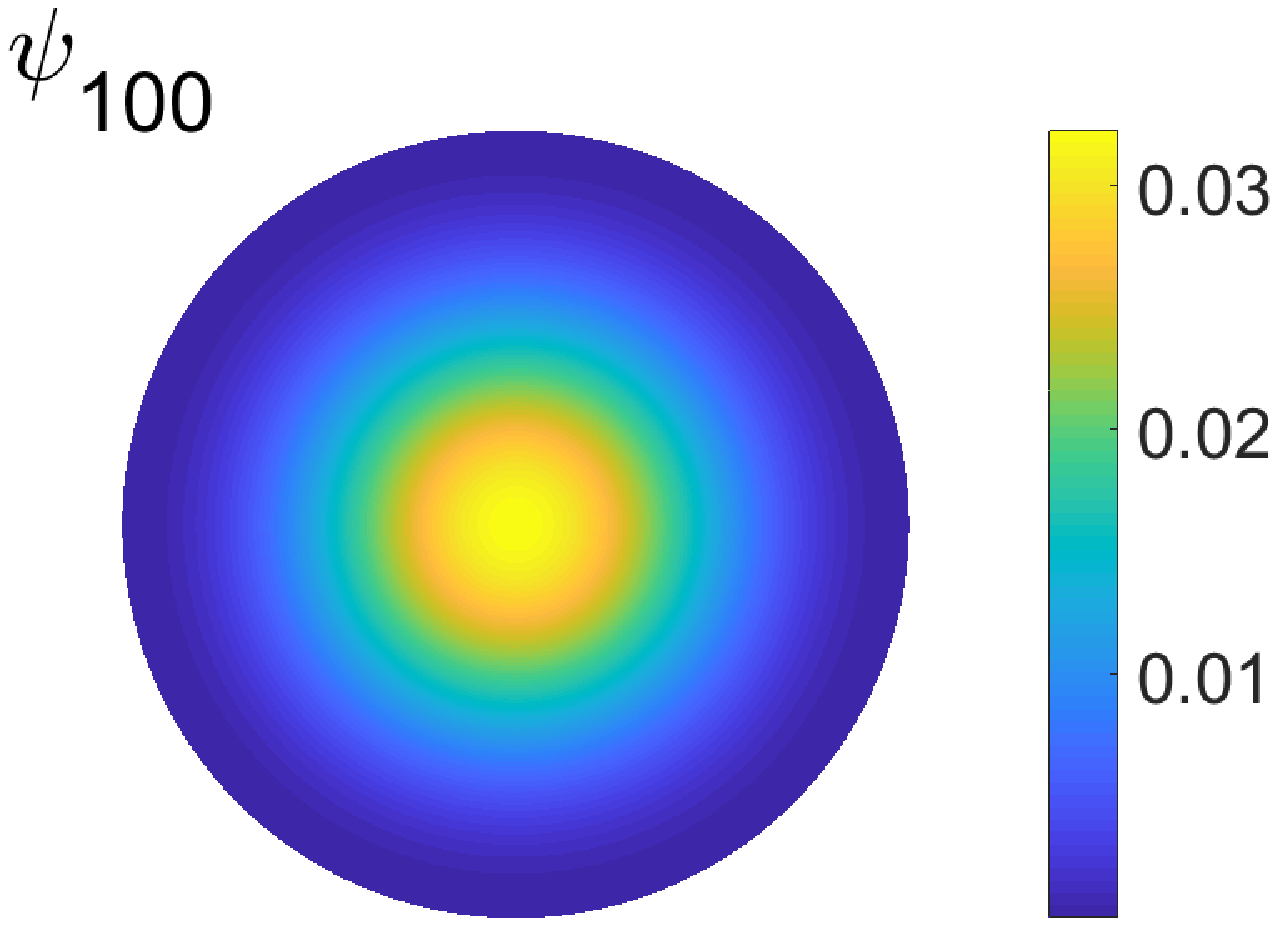}\\
	\includegraphics[width=.15\textwidth]{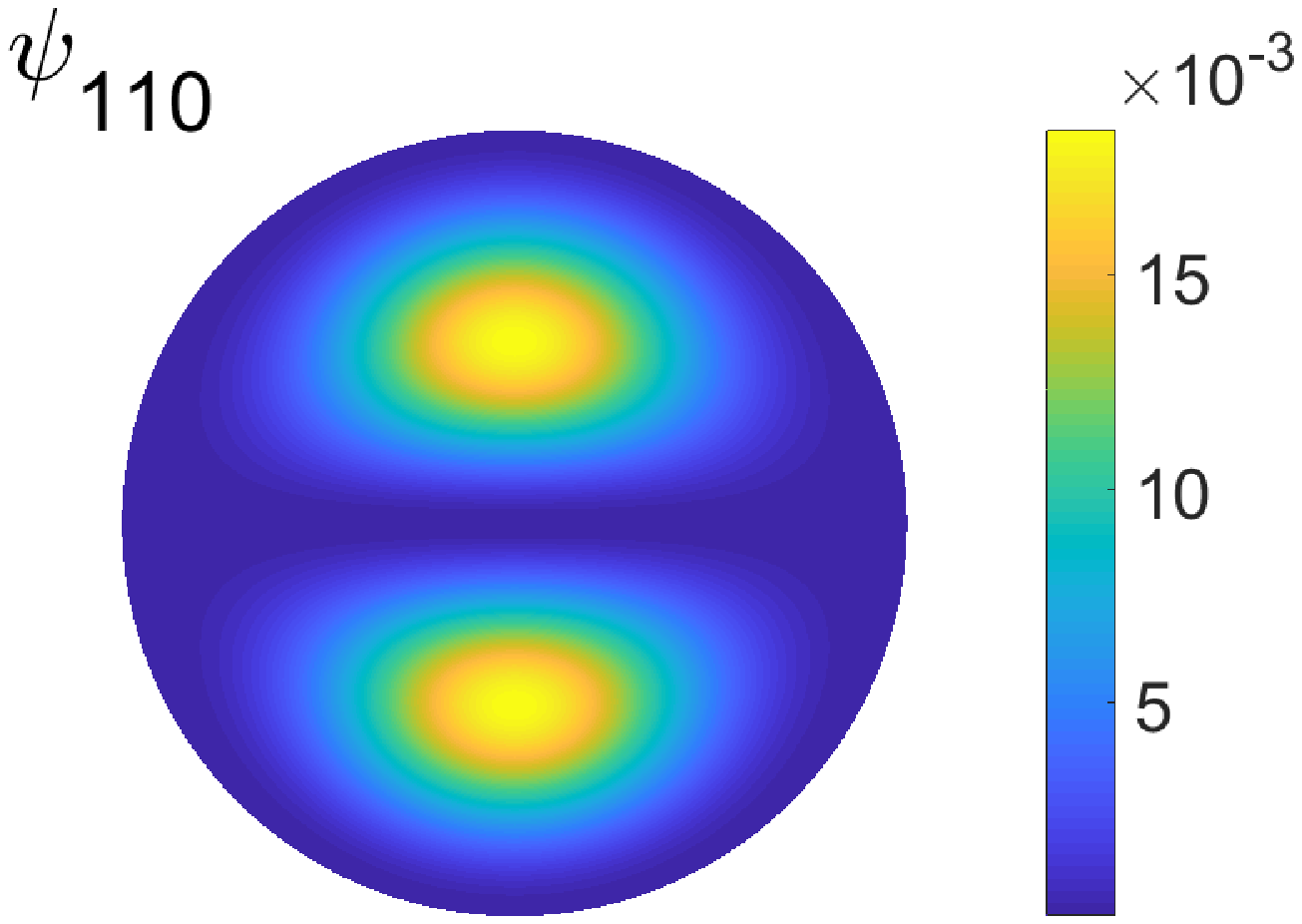}
	\includegraphics[width=.15\textwidth]{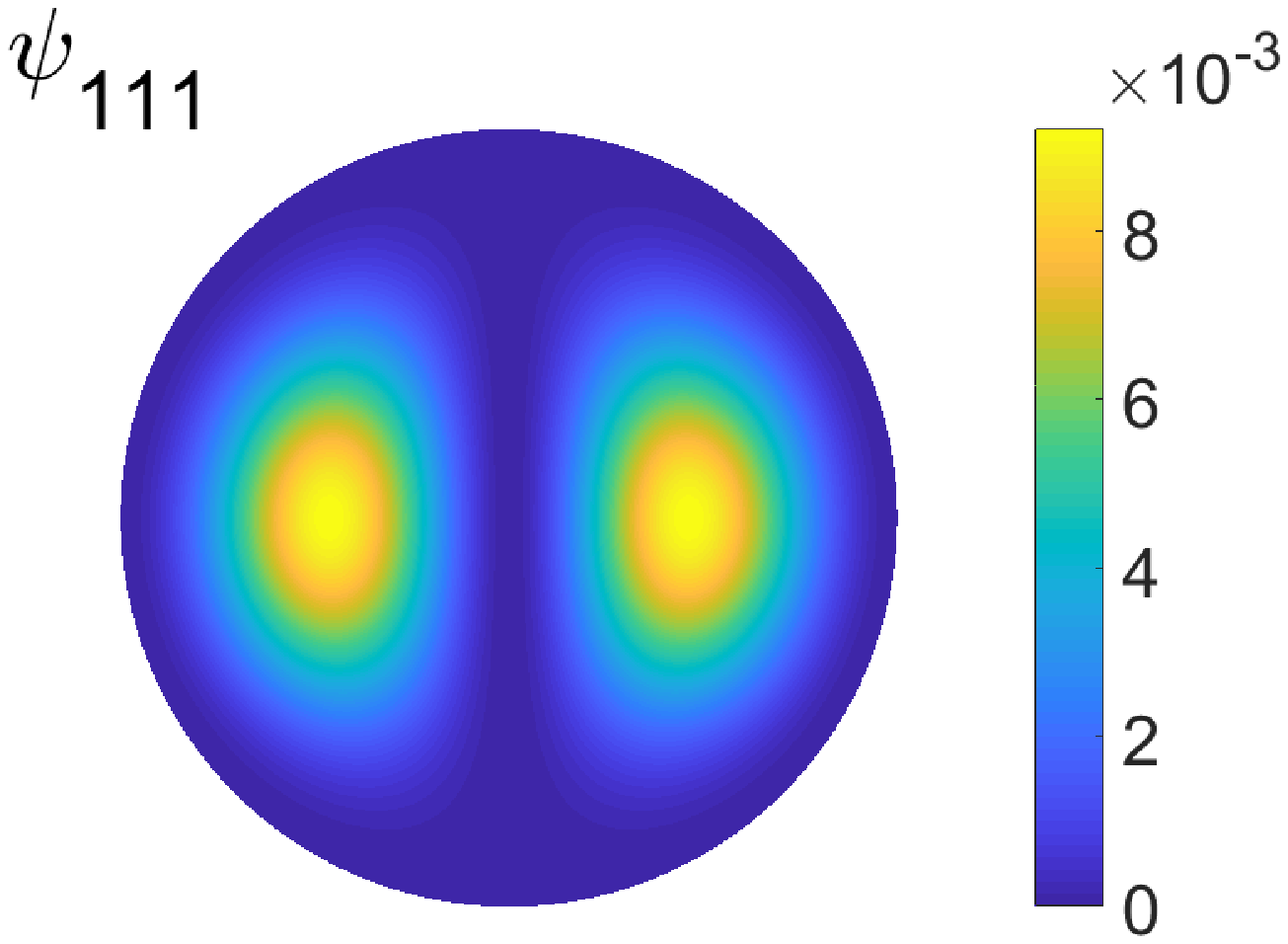}\\
	\includegraphics[width=.15\textwidth]{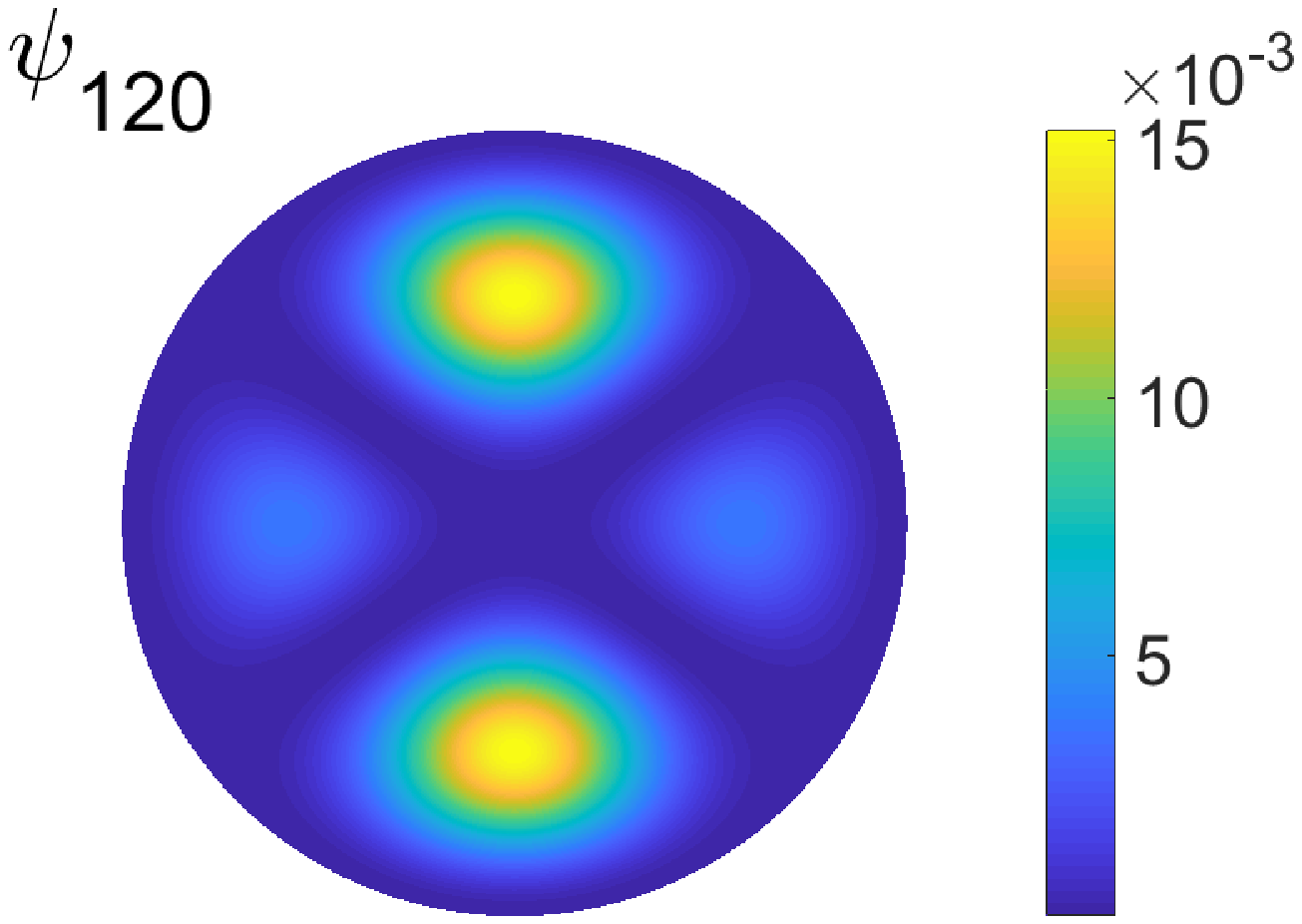}
	\includegraphics[width=.15\textwidth]{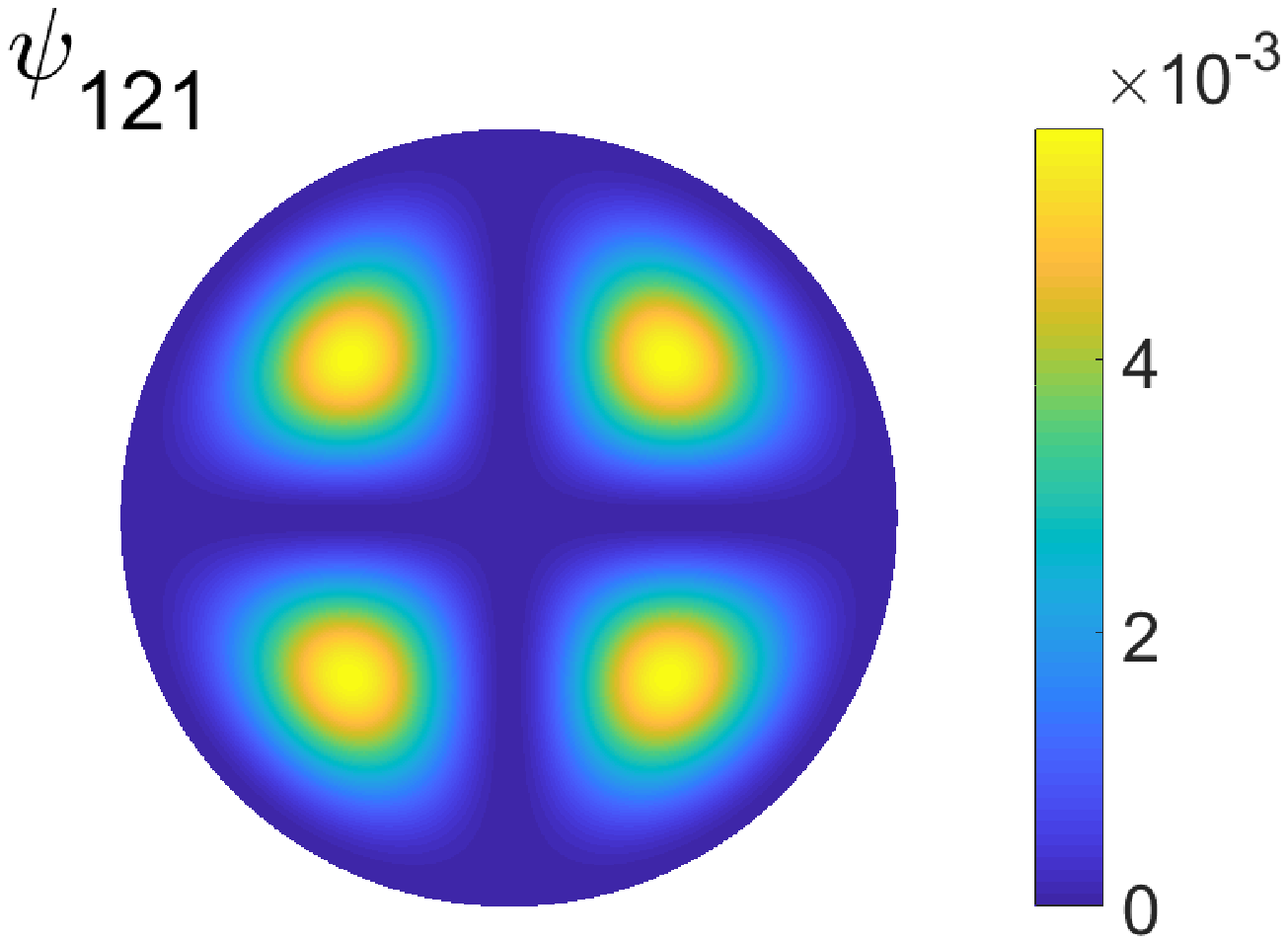}
	\includegraphics[width=.15\textwidth]{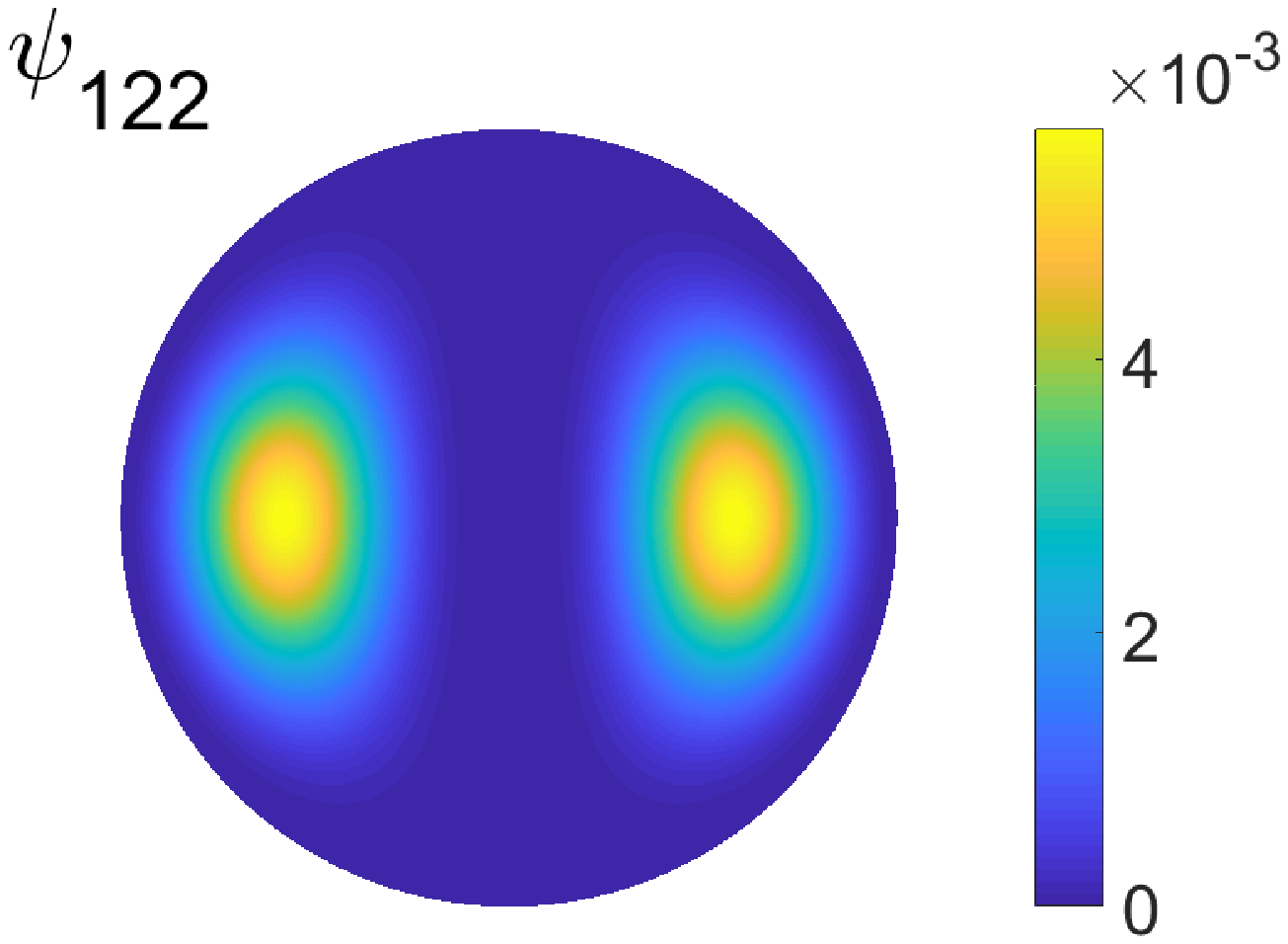}\\
	\includegraphics[width=.15\textwidth]{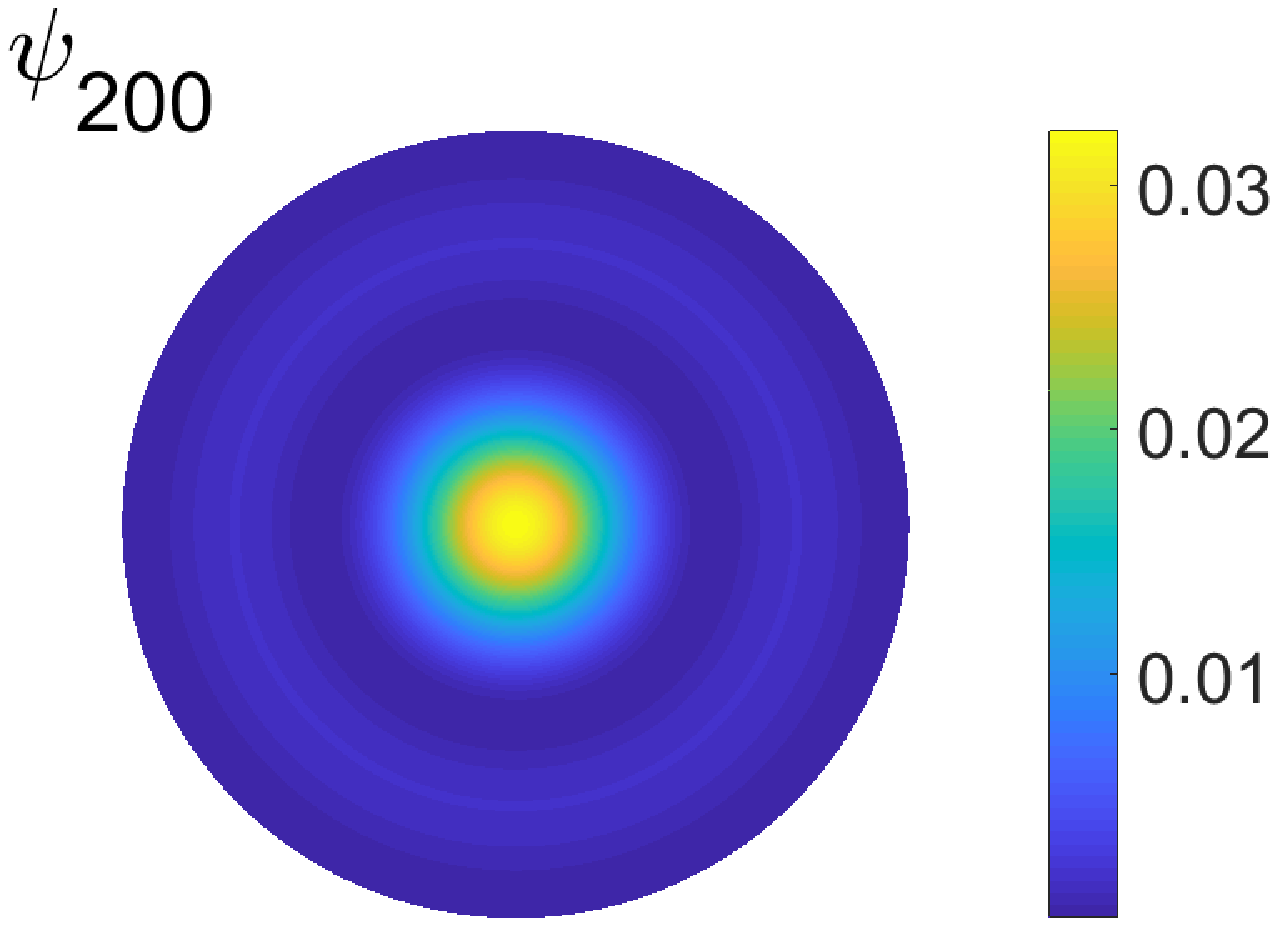}
	\caption{\label{fig:wfs}(Color online) The first four lowest order wavefunctions of QD. All modes have $2n+1$ degeneracy. Each row represents degenerate modes for $s\ge 0$.}	
\end{figure}

In this paper, the active medium is a QD array. It is assumed that the effects of QDs on each other is negligible, so this section concentrates on the individual QDs. The wavefunctions and eigenenergies should be derived by solving the well known Schr\"{o}dinger equation. Due to the spherical symmetry of the potential, wavefunctions are similar to angular momentum operator eigenfunctions. For the sake of simplicity and getting an insight to the whole problem, Schr\"{o}dinger equation is solved analytically using the infinite wall boundary conditions,
\begin{equation}
\psi_{kns}(r,\theta,\phi)=\left\lbrace\begin{array}{ll}
A_{nk}j_n\left(\frac{x_{nk}}{b}r\right)\yr_n^s(\theta,\phi) &r\le b,\\
0 &r>b,
\end{array}\right.
\end{equation}
where $x_{nk}$ is the $k$'th zero of $n$'th order spherical Bessel function of the first kind, $j_n$, and $\yr_n^s$'s are spherical harmonics. In the above relation $A_{nk}$ is a normalization constant which equals to
\begin{equation}
	A_{nk}=\left(\frac{2}{b^3[j_{n+1}(x_{nk})]^2}\right)^{0.5}.
\end{equation}
All the modes have $2n+1$ degeneracies. The eigenenergy of the $kns$'th mode is given by
\begin{equation}
E_{kns}=\frac{\hbar^2x_{nk}^2}{2m^*_\mathrm{QD}b^2},
\end{equation}
where $m^*_\mathrm{QD}$ is electron's effective mass of the material used for QD's construction. Some lowest order modes are sketched in Figure~\ref{fig:wfs} for $\phi=0$. 

The most important quantity in the gain medium which is required for the ongoing sections is the dipole moment. Quantum mechanical version of dipole moment between two states $\left|p\right>=\left|kns\right>$ and $\left|q\right>=\left|k'n's'\right>$ is defined as $\mathbf{d}_{pq}=-e\left\langle kns\left|r\right|k'n's'\right\rangle$. After integration and some manipulations, the following result is obtained,
\begin{equation}
\mathbf{d}_{pq}=-\hat{\mathbf{r}}\delta_{nn'}\delta_{ss'}ebf_{nkk'},\label{eq:acdp}
\end{equation}
where $f$ is a dimensionless parameter which is independent of the choice of geometry,
\begin{equation}
f_{nkk'}=\frac{2}{j_{n+1}(x_{nk})j_{n+1}(x_{nk'})}\int_0^1r^3j_n(x_{nk}r)j_{n}(x_{nk'}r)\,\mathrm{d}r.
\end{equation}
Considering Eq.~(\ref{eq:acdp}), it is seen that dipole moments have only radial components; It is clear that nonzero dipole moments exist between states with the same angular indices, $ns$.

A note should be stated here; In this section, the dipole moment of a single QD in the array is calculated in its rest reference frame. But we are going to use this value for all QDs in the array and for all reference frames in the next section. This statement is true because dipole moment is coordinate independent if the total charge is zero which is in our case.

QD should have three main energy levels. These states are called ground, the first, and second excited states. The energy difference between the second excited and ground states should be equal to photon's energy of the pump field which is an external laser source. And the difference between the first excited and ground states should be designed to resonance with the LSP mode of interest.

From now on, it is assumed that only a single mode of LSPs, $l=L$, is nearly in resonance with QD's electron transitions. The resonant transition in QDs is assumed to be $p\rightarrow q$ with eigenenergies $E_p$ and $E_q$, respectively. The radial component of this transition's dipole moment is denoted by $d$. Using this assumption, The active Hamiltonian can be written in the following form,\cite{scully1997quantum}
\begin{equation}
	H_\mathrm{g}=\frac{\hbar\omega_{qp}}{2}\hat{\sigma}_z,
\end{equation}
where
\begin{eqnarray}
\hat{\sigma}_z &=& \left|q\right>\!\left<q\right|-\left|p\right>\!\left<p\right|,\\
\hbar\omega_{qp} &=& E_q-E_p.
%&&\hat{\sigma}_+=\left|q\right>\!\left<p\right|,\\ &&\hat{\sigma}_-=\left|p\right>\!\left<q\right|,
\end{eqnarray}
In the above relations, we assume $E_q>E_p$, without loss of generality.
%\begin{eqnarray}
%\hat{H} &=&\sum_m\hbar\omega_L\hat{a}_{Lm}^\dagger\hat{a}_{Lm}+\frac{\hbar\omega_{qp}}{2}\hat{\sigma}_z\nonumber\\
%&-&\sum_{nm}\hbar\left(\Omega_{nm}^*\hat{a}_{Lm}^\dagger\hat{\sigma}_-+\Omega_{nm}\hat{\sigma}_+\hat{a}_{Lm}\right).
%\end{eqnarray}

\section{\label{sec:IntHam}Interaction Hamiltonian and spasing}
Using the definition of Rabi frequency, $\Omega=-\mathbf{E}\cdot\mathbf{d}/\hbar$,\cite{scully1997quantum} interaction Hamiltonian can be written as follows,
\begin{equation}
	H_\mathrm{I}=-\sum_{nm}\hbar\left(\Omega_{nm}^*\hat{a}_{Lm}^\dagger\hat{\sigma}_-+\Omega_{nm}\hat{\sigma}_+\hat{a}_{Lm}\right),
\end{equation}
where the azimuthal index takes the values of $m=-L, \cdots,L$ and $n$ is a dummy variable which labels the $n$'th QD in the array and runs over $1, \cdots,N$, where $N$ is the total number of QDs. In the above relation, $\Omega_{nm}$ is Rabi frequency corresponding to $n$'th QD and azimuthal index $m$. Rabi frequency is derived as follows,
\begin{equation}
\Omega_{nm}=-\frac{(L+1)\gamma_{L}d}{a\hbar}\,\yr_L^m(\theta_n,\phi_n),\label{eq:rabi}
\end{equation}
where $\theta_n$ and $\phi_n$ are the angular coordinates of $n$'th QD. Rabi frequency has the most vital role in the spasing condition,\cite{stockman2010spaser}
 \begin{equation}
 \frac{(\gamma'_L+\Gamma_{qp})^2}{(\gamma'_L+\Gamma_{qp})^2+(\omega_{qp}-\omega_L)^2}\sum_{nm}\left|\Omega_{nm}\right|^2\geq \gamma'_L\Gamma_{qp},\label{eq:spscond}
 \end{equation}
where $\Gamma_{qp}$ is the damping rate of polarization in QD. By substituting the Rabi frequency, Eq.~(\ref{eq:rabi}), into spasing condition, Eq.~(\ref{eq:spscond}), and assuming near resonance region, $\omega_{qp}\approx\omega_L$, after changing the summation over $n$ into integration and some other manipulations, we find that spasing occurs when the quality factor of LSP mode, $Q_L$, becomes higher than $Q_L^\mathrm{min}$,
 \begin{equation}
 Q_L^\mathrm{min}=\frac{\hbar\epsilon_0a^3\Gamma_{qp}[L\epsilon_\mathrm{rp}+(L+1)\epsilon_\mathrm{ra}]}{(L+1)^2|d|^2I_L}.\label{eq:qmin}
 \end{equation}
In the above relation, $I_L$ is defined as follows,
\begin{equation}
	I_L=\sum_{m=-L}^L\int_{4\pi}\varrho_\Omega(\theta,\phi)\left|\yr_L^m(\theta,\phi)\right|^2\,\mathrm{d}\Omega,\label{eq:IL}
\end{equation}
where $\varrho_\Omega$ is the number of QDs per unit solid angle with the dimension of inverse steradian (sr$^{-1}$). It can be shown that for uniform distribution of QDs on sphere the following result is true,
\begin{equation}
	I_L=(2L+1)\varrho_\Omega.\label{eq:ILuni}
\end{equation}
A noteworthy result can be considered from Eq.~(\ref{eq:qmin}). The minimum quality factor does not depend on Fermi energy while the quality factor of LSP modes apparently is affected by $E_\mathrm{F}$. This means that we can tune the spaser by changing $E_\mathrm{F}$. It is useful to define the threshold Fermi energy for mode $L$ that for Fermi energies larger than that the spasing occurs. This threshold could be derived by comparing Eq.~(\ref{eq:qmin}) and Eq.~(\ref{apxq}). Doing so yields the following result,
\begin{equation}
	E_\mathrm{F,th}^L=\frac{\pi\hbar^4\epsilon_0^3}{e^6\tau^2}\times\frac{\Gamma_{qp}^2}{f_{qp}^4}\times\frac{a^7}{b^4}\times\frac{[(L+1)\epsilon_\mathrm{ra}+L\epsilon_\mathrm{rp}]^3}{L(L+1)^5I_L^2}.
\end{equation}
Fermi level can be tuned by several methods including chemical doping and electrostatic gating. 
%In this paper the socond method is used. The gating voltage, $V$, is applied between graphene sphere and sheet. So threshold voltage is obtained as the following,
%\begin{equation}
%	V_L^\mathrm{th}=\frac{eE_\mathrm{F,th}^{L^2}}{\pi\hbar^2v_\mathrm{F}^2C},
%\end{equation}
%where $C$ is capacitance of the structure per unit area of graphene sphere that for $t$ indicating the minimum distance between graphene sphere and sheet is equal to
%\begin{equation}
%	C=\frac{4\epsilon_0\epsilon_\mathrm{ra}}{\pi a}\sum_{n=1}^\infty \frac{\sinh[\ln(t/a)+\sqrt{t^2/a^2-1}]}{\sinh[n\ln(t/a)+\sqrt{t^2/a^2-1}]}.
%\end{equation}

%\section{\label{sec:coupgrsheet}Coupling to Graphene Sheet}

The near field of our proposed spaser could be used for exciting SPs on flat interfaces such as graphene sheets or metal films. The near field has a wide range of wavevectors which one of them could be phase-matched with a SP and excites it.
%\vspace{.8cm}
  
\section{conclusion}
In summary, we proposed a new structure for spasing consisting of graphene. The proposed structure is made up of a graphene nanosphere and an array of QDs. The QD array plays the role of gain medium. The spaser has been thoroughly analyzed theoretically by using full quantum mechanical description. After analyzing the spaser, a necessary condition for spasing was derived. we found that spasing could occur when quality factor of some LSP mode becomes higher than some minimum value. Furthermore we translated the condition for quality factor to a criterion for Fermi energy and showed that by tuning the Fermi energy, one can select which LSP mode to spase.

%We suggested using of the near field of this spaser for exciting SPs on an interface, which in this paper is a graphene sheet.
%Through the paper, it was assumed that quasielectrostatic approximation is sufficiently accurate for analyzing the structure. This approximation was validated by comparing the derived polarizability of graphene nanosphere with the results of the numerical solution of full wave maxwell's equations.

%\bibliography{references}
%\input{mypaper.bbl}

\end{document}